\pdfoutput=1
\documentclass{elsart5p}
\usepackage{amsmath, amscd, amssymb}
\usepackage{graphicx}
\usepackage[small, bf]{caption}
\usepackage{enumerate}
\usepackage{paralist}
\usepackage{multirow}
\usepackage{array}
\usepackage{rotating}
\usepackage{enumitem}
\usepackage[all, import]{xy}
\usepackage{multirow}
\usepackage{cite}

\graphicspath{{figs/dlnyhdg/}}
\def\R{\mathbb{R}}

\def\Diff_vol{\textrm{Diff$_\textrm{vol}$}}

\DeclareMathOperator{\grad}{grad}

\def\div{\operatorname{div}}

\DeclareMathOperator{\hodge}{\ast} 
\DeclareMathOperator{\dual}{\star}

\long\def\symbolfootnote[#1]#2{\begingroup%
\def\thefootnote{\fnsymbol{footnote}}\footnote[#1]{#2}\endgroup}

\newenvironment{xyoverpic}[3]
{%
\begin{xy}
\xyimport#1{\includegraphics[#2]{#3}}
}{\end{xy}}

\newenvironment{cxyoverpic}[3]
{%
\begin{center}
\centering\leavevmode\small
\begin{xyoverpic}{#1}{#2}{#3}
}{\end{xyoverpic}
\end{center}}

\begin{document}

\begin{frontmatter}

\vspace{-1.5in}
\title{Delaunay Hodge Star}

\author[addr:ui]{Anil N.~Hirani\corauthref{athr:corr}},
\ead{hirani@illinois.edu}
\author[addr:rpi]{Kaushik Kalyanaraman},
and \author[addr:anl]{Evan B.~VanderZee}

\corauth[athr:corr]{Corresponding author}

\address[addr:ui]{Department of Mathematics, University of Illinois at
  Urbana-Champaign, 1409 W. Green St., Urbana, IL}
\address[addr:rpi]{SCOREC, Rensselaer Polytechnic Institute, 110 8th
  St., Troy, NY}
\address[addr:anl]{Argonne National Laboratory, 9700 S. Cass Ave.,
  Lemont, IL}

\begin{abstract}
  We define signed dual volumes at all dimensions for circumcentric
  dual meshes. We show that for pairwise Delaunay triangulations with
  mild boundary assumptions these signed dual volumes are
  positive. This allows the use of such Delaunay meshes for Discrete
  Exterior Calculus (DEC) because the discrete Hodge star operator can
  now be correctly defined for such meshes. This operator is crucial
  for DEC and is a diagonal matrix with the ratio of primal and dual
  volumes along the diagonal. A correct definition requires that all
  entries be positive. DEC is a framework for numerically solving
  differential equations on meshes and for geometry processing tasks
  and has had considerable impact in computer graphics and scientific
  computing. Our result allows the use of DEC with a much larger class
  of meshes than was previously considered possible.

\end{abstract}


\end{frontmatter}

\noindent \textbf{NOTE: } This version incorporates corrections to the
previous arXiv version (v3). The previous version (minus the Appendix)
appeared as the journal article~\cite{HiKaVa2013}. % All section and
% figure numbers in this version, previous arXiv version v3 and
% in~\cite{HiKaVa2013} are identical. 
The main correction in this version is replacement of Figure~1
of~\cite{HiKaVa2013} by the Figure~\ref{fig:poissons} here. The change
is in columns 3 and 4 of the figure. This error in~\cite{HiKaVa2013}
(and hence in arXiv version v3 of this article) has been traced to a
programming error resulting from the generation and processing of the
meshes used in columns 3 and 4 of Figure~1. This error resulted in
incorrect boundary conditions being used for the meshes in columns 3
and 4 and misidentification of some edges in column 4. We provide
additional details in Section~\ref{sec:corrigendum} in the Appendix of
this arXiv version.  The implication of this correction of
Figure~\ref{fig:poissons} is the possibility it raises that Discrete
Exterior Calculus may be applicable in an even broader class of meshes
than was believed possible when the previous version appeared. This
version also fixes a formula error in Section~2 and error with an
illustration in Figure~6 of~\cite{HiKaVa2013}. There are no errors in
the mathematical statements (lemmas and theorems)
of~\cite{HiKaVa2013}.

\section{Introduction} \label{sec:intr}

Discrete Exterior Calculus (DEC) is a framework for numerical solution
of partial differential equations on simplicial meshes and for
geometry processing tasks~\cite{Hirani2003short,DeKaTo2008short}. DEC
has had considerable impact in computer graphics and scientific
computing. It is related to finite element exterior calculus and
differs from it in how inner products are defined. The main objects in
DEC are $p$-cochains, which for the purpose of this paper may be
considered to be a vector of real values with one entry for each
$p$-dimensional simplex in the mesh. For $p$-cochains $a$ and $b$
their inner product in DEC is $a^T \hodge_p b$ where $\hodge_p$ is a
diagonal discrete Hodge star operator. This is a diagonal matrix of
order equal to the number of $p$-simplices and with entries that are
ratios of volumes of $p$-simplices and their $(n-p)$-dimensional
circumcentric dual cells.

\begin{figure*}[t]
  \centering
 \begin{tabular}{cccc}
    & Unsigned & Bad Boundary & Not Delaunay \\
    \includegraphics[scale=0.24, trim=0.0in 0.0in 0.0in 0.0in, clip]
    {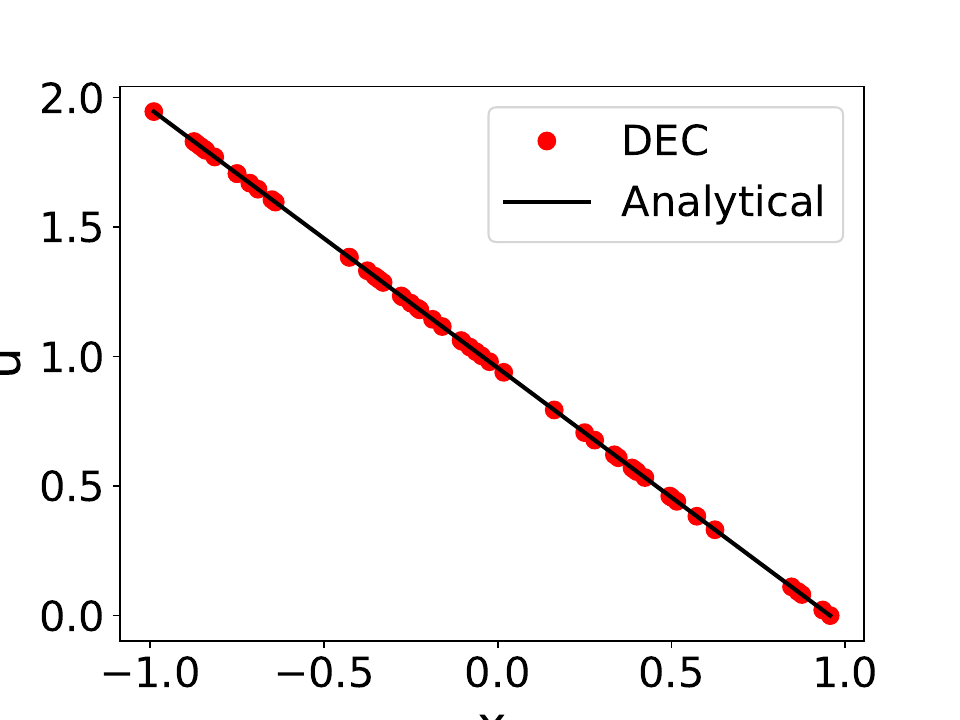} &
    \includegraphics[scale=0.24, trim=0.0in 0.0in 0.0in 0.0in, clip]
    {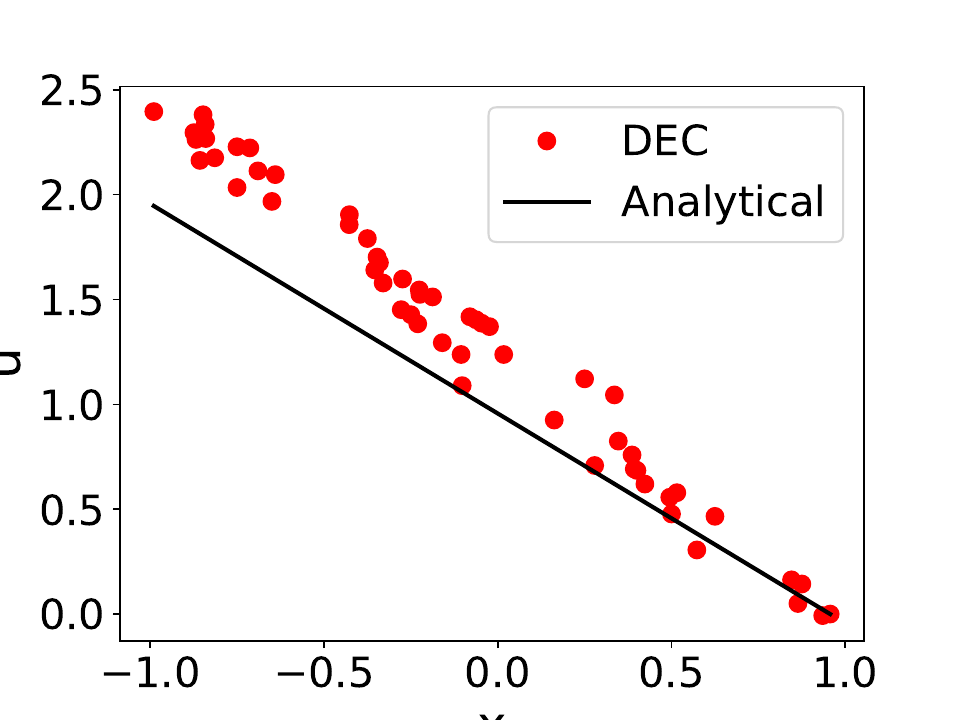}
    & 
    \includegraphics[scale=0.24, trim=0.0in 0.0in 0.0in 0.0in, clip]
    {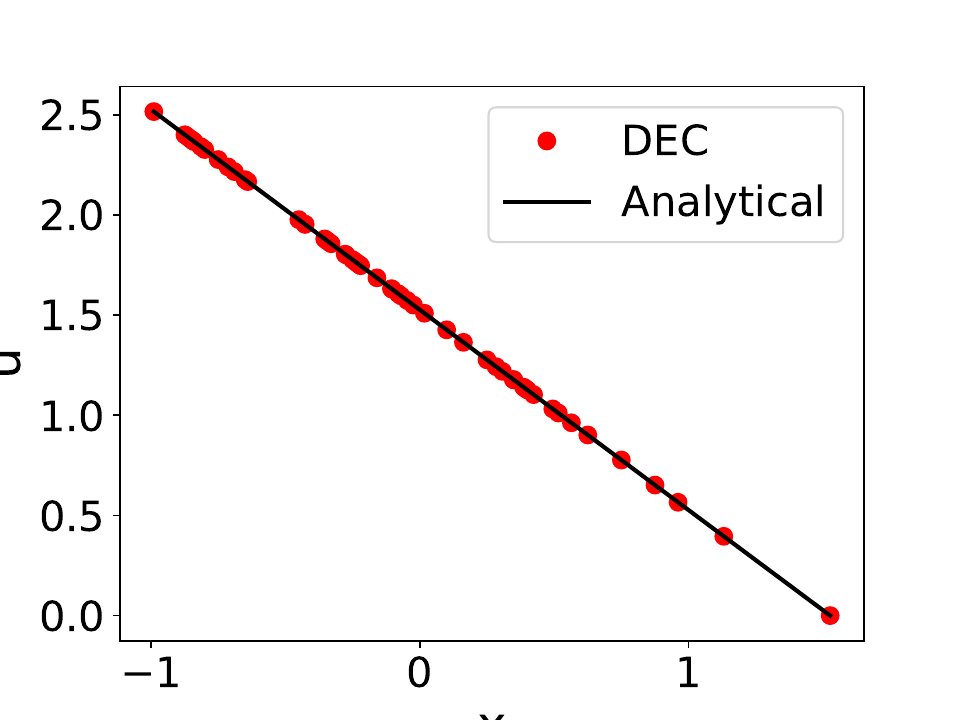} &
    \includegraphics[scale=0.24, trim=0.0in 0.0in 0.0in 0.0in, clip]
    {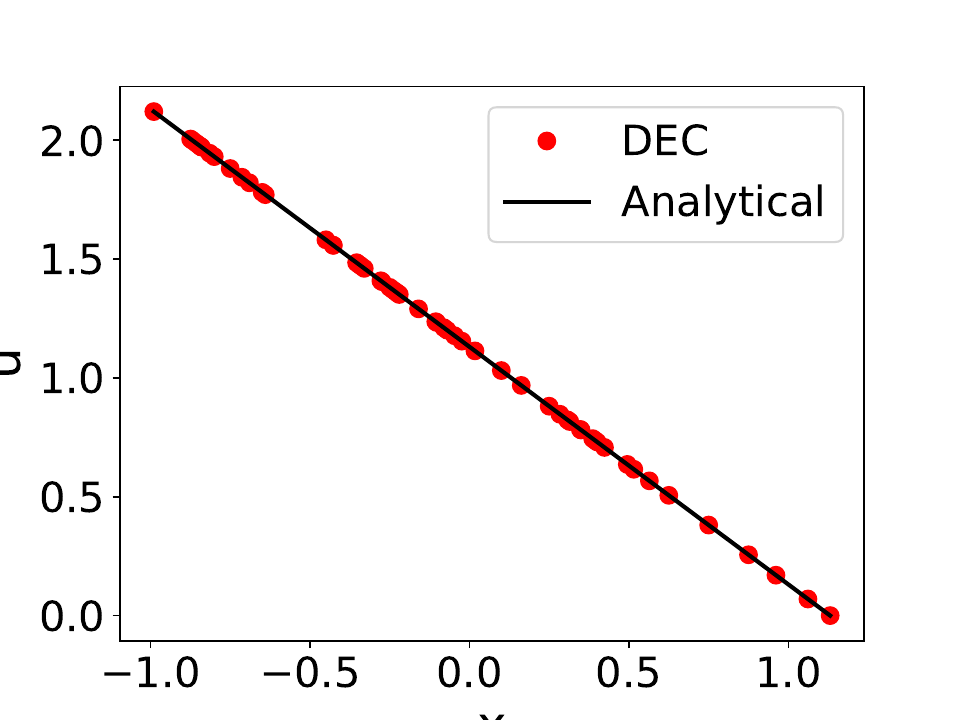} \\
    \includegraphics[scale=0.35, trim=1.35in 0.5in 1.35in 0.5in, clip]
    {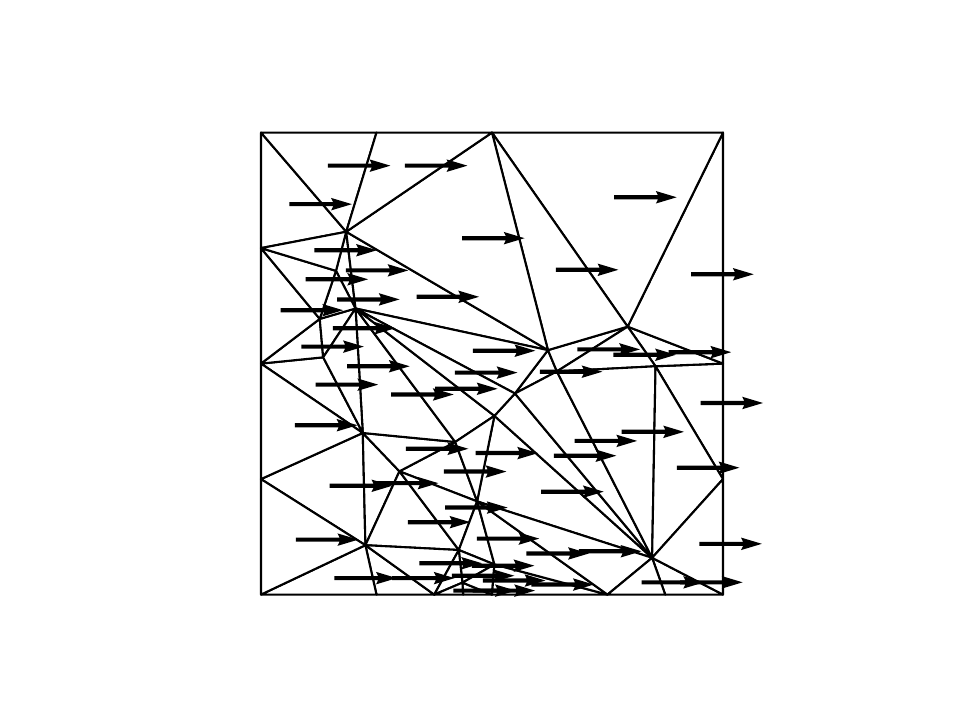} &
    \includegraphics[scale=0.35, trim=1.35in 0.5in 1.35in 0.5in, clip]
    {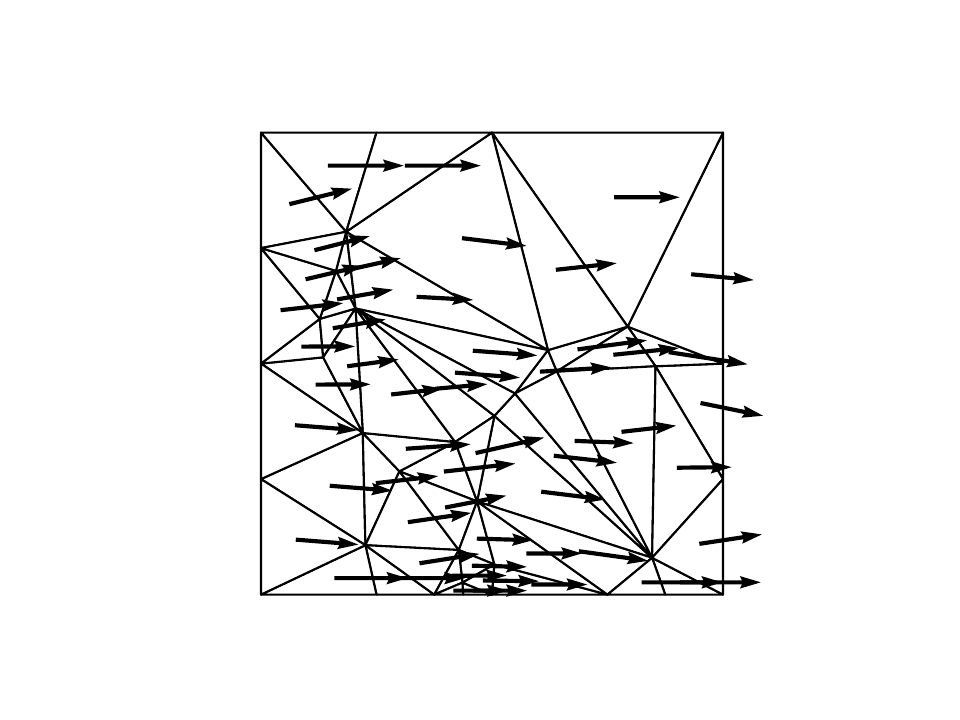}
    & 
    \includegraphics[scale=0.35, trim=1.35in 0.5in 1.35in 0.5in, clip]
    {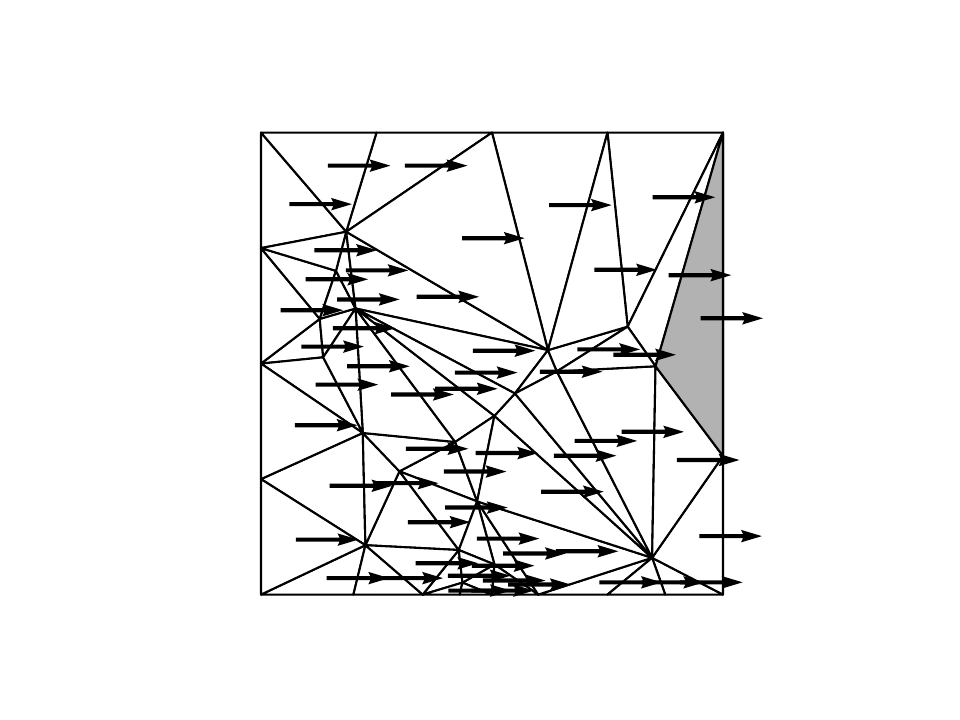} &
    \includegraphics[scale=0.35, trim=1.35in 0.5in 1.35in 0.5in, clip]
    {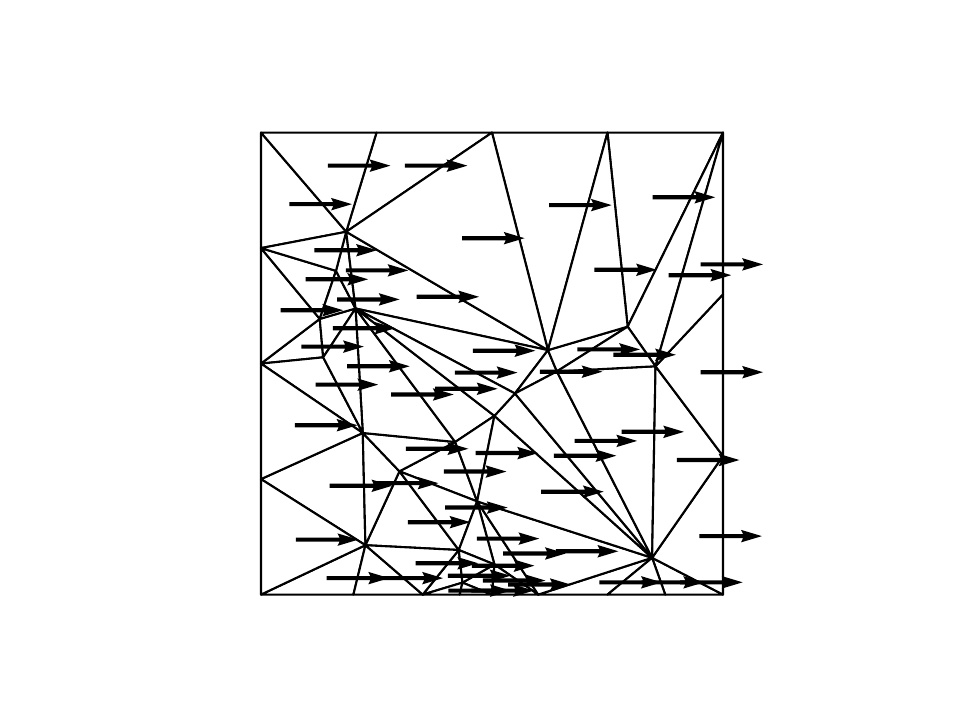}
  \end{tabular}
  \caption{Solution of Poisson's equation $-\Delta u = f$ in mixed
    form. In mixed form this equation is the system
    $\sigma = -\grad u$ and $\div \sigma = f$. The boundary condition
    is constant influx on left and outflux on right. The correct
    solution is linear $u$ which varies only along x-direction and a
    constant horizontal $\sigma$. The top row shows $u$ and bottom row
    shows $\sigma$. The first column shows the correct solution using
    the results of this paper on a Delaunay mesh with correct boundary
    simplices. The second column is for unsigned duals using the same
    mesh as first column. This fails to produce the correct
    solution. The next two columns are instances in which the
    $1$-Hodge star mass matrix is not positive definite. The third
    column has a single bad (i.e., not one-sided) boundary triangle
    shown shaded in a Delaunay mesh. The fourth column is a non
    Delaunay mesh. It appears that Discrete Exterior Calculus produces
    correct solution even for these cases for meshes used in columns 3
    and 4.}
  \label{fig:poissons}
\end{figure*}

For this to define a genuine inner product the entries have to be
positive. Simply taking absolute values or considering all volumes to
be unsigned does not lead to correct solutions of partial differential
equations. See Figure~\ref{fig:poissons} to see the spectacular failure
when unsigned volumes are used in solving Poisson's equation in mixed
form.

When DEC was invented it was known that completely well-centered
meshes were sufficient but perhaps not necessary for defining the
Hodge star operator. (A completely well-centered mesh is one in which
the circumcenters are contained within the corresponding simplices at
all dimensions. Examples are acute-angled triangle meshes and
tetrahedral meshes in which each triangle is acute and each
tetrahedron contains its circumcenter.) For such meshes, the volumes
of circumcentric dual cells are obviously positive. For some years now
there has been numerical evidence that for codimension 1 duality some
(but not all) pairwise Delaunay meshes yield positive dual volumes if
volumes are given a sign based on some simple rules. Moreover, these
seem to yield correct numerical solutions for a simple partial
differential equation. A pairwise Delaunay mesh (of dimension $n$
embedded in $\R^N$, $N\ge n$) is one in which each pair of adjacent
$n$-simplices sharing a face of dimension $n-1$ is Delaunay when
embedded in $\R^n$. (Imagine a pair of triangles with a hinge at the
shared edge and lay the pair flat on a table.) This generalizes the
Delaunay condition to triangle mesh surfaces embedded in three
dimensions and analogous higher dimensional settings. For planar
triangle meshes and for tetrahedral meshes in three dimensions
pairwise Delaunay is same as Delaunay.

We give a sign convention for dual cells and a mild assumption on
boundary simplices. With these in hand, for pairwise Delaunay meshes
it is easy to see that the codimension 1 dual lengths are positive in
the most general case (dimension $n$ mesh embedded in $\R^N$). In
addition, we prove that such triangle meshes embedded in two or three
dimensions have positive vertex duals and that the duals of vertices
and edges of tetrahedral meshes in three dimensions are positive. This
settles the question of positivity for all dimensions of duality for
simplicial meshes used in physical and graphics applications and opens
up the possibility of using DEC with a much larger class of meshes.

Note that the results of this paper are relevant for the
\emph{assembly} of the duals from elementary duals (defined in
Section~\ref{sec:sgnddlclls}). Thus it is important for algorithms
which compute these dual volumes piece by piece from the elementary
duals (such as are used in the software PyDEC~\cite{BeHi2012}). It may
be possible to have alternative formulas which bypass the assembly
process and give the dual volumes directly. Such formulas are hinted
at in \cite{MuMeDeDe2011short}. Zobel~\cite{Zobel2010short} has also
hinted at the positivity of the duals in some cases but without proofs
or details.

There are alternatives to the discrete diagonal Hodge star. In finite
element exterior calculus, the role of discrete diagonal Hodge star is
played by the mass matrix corresponding to polynomial differential
forms~\cite{ArFaWi2010short}. This matrix is in general not
diagonal. Sometimes these are referred to as the Galerkin Hodge
stars. Our results do not apply to that case directly. For finite
elements, there are other quality measures that may be important. See
for instance~\cite{Shewchuk2002ashort}.

\medskip\noindent {\bf Notation: } We will sometimes write the
dimension of a simplex as a superscript. The notation $\tau \prec
\sigma$ means that $\tau$ is a face of $\sigma$. Circumcenter of a
simplex $\tau$ is denoted $c_\tau$ and the circumcentric dual of
$\tau$ as $\dual \tau$. Some figure labels like $\dual p$ stand for
dual of a $p$-simplex.

\section{Signed Circumcentric Dual Cells} \label{sec:sgnddlclls}

The dual mesh of a (primal) simplicial mesh is often defined using a
barycentric subdivision. For every $p$-dimensional simplex, there is a
dual $(n - p)$-dimensional cell where $n$ is the dimension of the
simplicial complex. In DEC circumcenters are used instead of
barycenters because the resulting orthogonality between the primal and
dual is an integral part of the definition of some of the operators of
DEC~\cite{Hirani2003short}. The circumcentric dual cell of a
$p$-dimensional primal simplex $\tau$ is constructed from a set of
simplices incident to the circumcenter of $\tau$. These are called
\emph{elementary dual simplices} with vertices being a sequence of
circumcenters of primal simplices incident to $\tau$. The sequence
begins with the circumcenter of $\tau$, moves through circumcenters of
higher-dimensional simplices $\sigma^i$ and ends with the circumcenter
of a top-dimensional simplex $\sigma^n$ such that $\tau \prec
\sigma^{p + 1} \prec \cdots \prec \sigma^n$. Taking each of the
possibilities for $\sigma^i$ at each dimension $i$ yields the full
dual cell.

In the past in the software PyDEC the volume of the dual cells has
been taken to be the sum of unsigned volumes of the elementary dual
simplices.  Instead we define the volume of the dual cells as the sum
of signed volumes of its elementary dual simplices. Our contribution
is in defining the sign convention and describing with proofs the
class of meshes for which the dual volumes and hence Hodge star
entries are positive. If the primal complex is completely well
centered every elementary dual has a positive volume.  In general, the
sign of the volume of an elementary dual simplex is defined as
follows. Start from the circumcenter of $\tau$.  Let $v_p$ be the
vertex such that $v_p * \tau$ is the simplex $\sigma^{p + 1}$ formed
by the vertices of $\tau$ together with $v_p$. Similarly, for $p + 1
\leq i \leq n - 1$, let $v_i$ be the vertex such that $v_i * \sigma^i$
is the simplex $\sigma^{i + 1}$. If the circumcenter of $\sigma^{p +
  1}$ is in the same half space of $\sigma^{p+1}$ as $v_p$ relative to
$\tau$, let $s_p = +1$, otherwise, $s_p = -1$. Likewise, for $p + 1
\leq i \leq n - 1$, if the circumcenter of $\sigma^{i + 1}$ is in the
same half space as $v_i$ relative to $\sigma^i$, let $s_i = +1$,
otherwise, $s_i = -1$.  Then, the sign $s$ of the elementary dual
simplex is the product of the signs at each dimension, that is, $s =
s_p \, s_{p + 1} \cdots s_{n-1}$.

For illustration of this sign rule, we now consider various cases in
two and three dimensions.  The first example is the dual of an edge
$ab$ in a triangle $abc$.  From the midpoint of $ab$ -- its
circumcenter -- we move to the circumcenter of triangle $abc$.  If
this move is towards vertex $c$, then the sign is $s = s_1 = +1$, but
if it is away from vertex $c$, as it will be if the angle at vertex
$c$ is obtuse, then the sign is $s = s_1 = -1$.  The next example is
the dual of vertex $a$ in triangle $abc$.  We will consider the
simplex formed from the circumcenter of $a$, the circumcenter of $ac$,
and the circumcenter of $abc$.  The move from $a$ to the midpoint of
$ac$ gives $s_0 = +1$, since vertex $c$ and the midpoint of $ac$ are
in the same direction from $a$.  The move from the midpoint of $ac$ to
the circumcenter of $abc$ gives $s_1 = +1$ if we go towards $b$ and
$s_1 = -1$ if we move away from $b$.  The sign of the volume of this
contribution to the dual of vertex $a$ is $s = s_0 \, s_1 = s_1$.

For a tetrahedron $abcd$ we can expand on the cases for triangle
$abc$.  For the dual to face $abc$, we move from the circumcenter of
$abc$ to the circumcenter of $abcd$.  If the circumcenter of $abcd$ is
in the same half space as vertex $d$ relative to $abc$, this move is
towards $d$, the sign is $s = s_2 = +1$, and the signed length
(volume) is positive; otherwise, it is negative.  Of the two
contributions to the dual of edge $ab$, we focus on the simplex formed
from the circumcenter of $ab$, the circumcenter of $abc$ and the
circumcenter of $abcd$.  The sign $s_1$ is determined as it was for
the dual of edge $ab$ in triangle $abc$.  The sign $s_2$ is $+1$ if
vertex $d$ and the circumcenter of $abcd$ are in the same half space
relative to $abc$.  Thus for the dual of edge $ab$, the sign of the
volume is $s = s_1 \, s_2$, and both $s_1$ and $s_2$ can be either
positive or negative.  As a final example, consider the simplex formed
from vertex $a$, the circumcenter of $ac$, the circumcenter of $abc$,
and the circumcenter of $abcd$.  This simplex contributes to the dual
of vertex $a$.  Signs $s_0$ and $s_1$ are the same as they were for
the dual of vertex $a$ in triangle $abc$.  Sign $s_2$ is $-1$ if
triangle $abc$ separates vertex $d$ from the circumcenter of
tetrahedron $abcd$.  The sign of this elementary volume then is $s =
s_0 \, s_1 \, s_2$.

\begin{figure}[h]
  \centering
  \begin{tabular}{cm{0.1in}cm{0.1in}c}
    & & $\star 1$ & & $\star 0$ \\
    \centering \begin{sideways} \centering \hspace{0.5in} Plus \end{sideways} & &
    \includegraphics[scale=0.4, trim=4.0in 1.5in 1.5in 1.5in, clip]
    {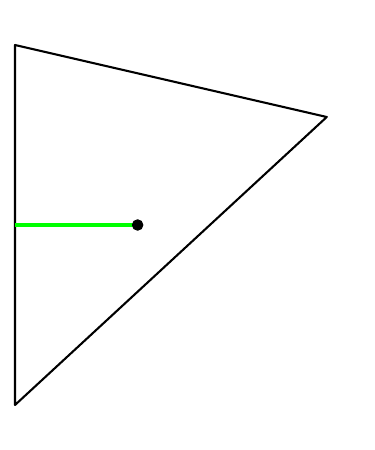} & &
    \includegraphics[scale=0.4, trim=4.0in 1.5in 1.5in 1.5in, clip]
    {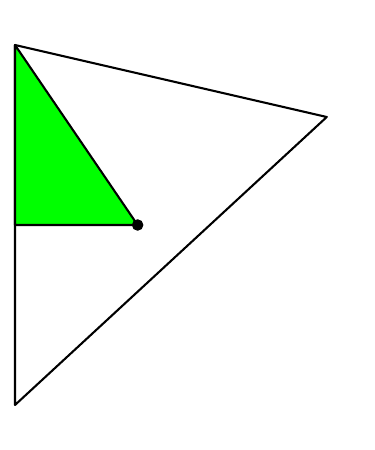} \\
    \centering \begin{sideways} \centering \hspace{0.5in} Minus \end{sideways} & &
    \includegraphics[scale=0.4, trim=3.4in 1.5in 3.0in 1.5in, clip]
    {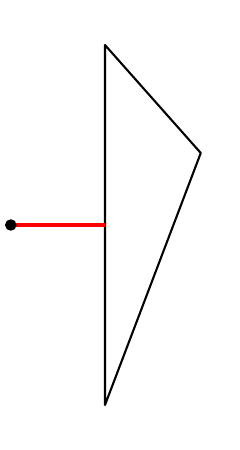} & &
    \includegraphics[scale=0.4, trim=3.4in 1.5in 3.0in 1.5in, clip]
    {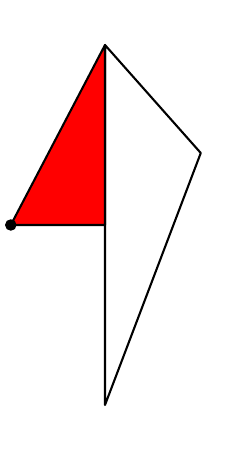}
  \end{tabular}
  \caption{Examples of sign rule application in 2d. The dot marks the
    circumcenter and green and red are used to denote positive and
    negative volumes respectively.}
  \label{fig:sgnxmpls2d}
\end{figure}

The significance of the sign rule defined above is that it orients the
elementary dual simplices in a particular way with respect to the dual
orientation for a completely well-centered simplex. Consider two
$n$-dimensional simplices $\sigma$ and $\sigma_w$ which have the same
orientation but such that $\sigma_w$ is well-centered. We are given a
bijection between the vertices of these two simplices such that the
resulting simplicial map is orientation preserving. This vertex map
induces a bijection between faces of the two simplices and between
their elementary duals. Let $\tau$ and $\tau_w$ be two corresponding
$p$-dimensional faces in the two simplices and consider their duals
$\dual \tau$ and $\dual \tau_w$. If we consider two corresponding
elementary duals in $\dual \tau$ and $\dual \tau_w$ we can affinely
map these such that the first vertex (the circumcenter of $\tau$ or
$\tau_w$) is mapped to the origin and the others are mapped to $+1$ or
$-1$ along a coordinate axis. For the elementary dual in $\dual
\tau_w$ we always choose $+1$ for all $n-p$ coordinate axes. For
$\dual \tau$ we choose $+1$ if the sign along that direction of the
elementary dual is positive according to the sign rule described above
and $-1$ otherwise. It is clear (and is easy to show using
determinants) that the orientation of the corresponding elementary
duals will be same if an even number of $-1$ directions are used for
the elementary dual in $\dual \tau$ and the orientations will be
opposite otherwise. Thus we have shown the following result.
\begin{thm}
  With $\sigma$, $\sigma_w$, $\tau$, and $\tau_w$ as above, the
  orientation of $\dual \tau$ is same as that of $\dual \tau_w$ if an
  even number of $-1$ signs appear according to sign rule and is
  opposite otherwise.
\end{thm}
If the orientation of $\dual \tau$ is same as $\dual \tau_w$ we will
assign a positive volume to $\dual \tau$ and otherwise a negative
volume.

\section{Signed Dual of a Delaunay Triangulation}
We first consider the codimension 1 case in the most general setting
of a simplicial complex of arbitrary dimension $n$ embedded in
dimension $N \ge n$.  After that we consider cases other than
codimension 1 but in more restricted settings. For these latter cases
we restrict ourselves to the physically most useful cases of triangle
meshes embedded in two or three dimensions ($n = 2$ and $N = 2$ or 3)
and tetrahedral meshes embedded in three dimensions ($n = N = 3$). We
conjecture that these results can be extended to the more general
setting of arbitrary $n$ and $N\ge n$ but those cases are not as
important for physical applications and we leave those for future
work. For the general codimension 1 case we first prove the following
basic fact about circumcenter ordering for Delaunay pairs.

\begin{lem}[Circumcenter Order]
 \label{lmm:crcmcntrsdlnypr} 
 Let $\tau$ be an $(n - 1)$-dimensional simplex in $\R^n$. Let $L$ and
 $R$ be points such that $\lambda = L * \tau$ and $\rho = R * \tau$
 form a non-degenerate Delaunay pair of $n$-dimensional simplices with
 circumcenters $c_{\lambda}$ and $c_{\rho}$, respectively. Then,
 $c_{\lambda}$ and $c_{\rho}$ have the same relative ordering with
 respect to $\tau$ as $L$ and $R$.
\end{lem}
\begin{pf}
  Consider the collection of $(n-1)$-dimensional spheres containing
  the vertices of $\tau$. Since $\lambda$ and $\rho$ are a
  non-degenerate Delaunay pair, their circumspheres are empty and
  belong to this collection. It is then easy to see that $c_{\lambda}$
  and $c_{\rho}$ will be in the same order as $\lambda$ and
  $\rho$. See Figure~\ref{fig:crcmcntrordr}.  For an alternative, more
  algebraic and detailed proof, see
  Appendix~\ref{appndx:crcmcntrordr}. (In fact there we show the
  stronger result that the correctness of the circumcenter ordering is
  equivalent to the simplices being a non-degenerate Delaunay pair.)
\end{pf}

\begin{figure}[h]
  \begin{cxyoverpic}
    {(1, 1)}{scale=0.7, trim=2.2in 1.9in 1.5in 1.3in, clip}
    {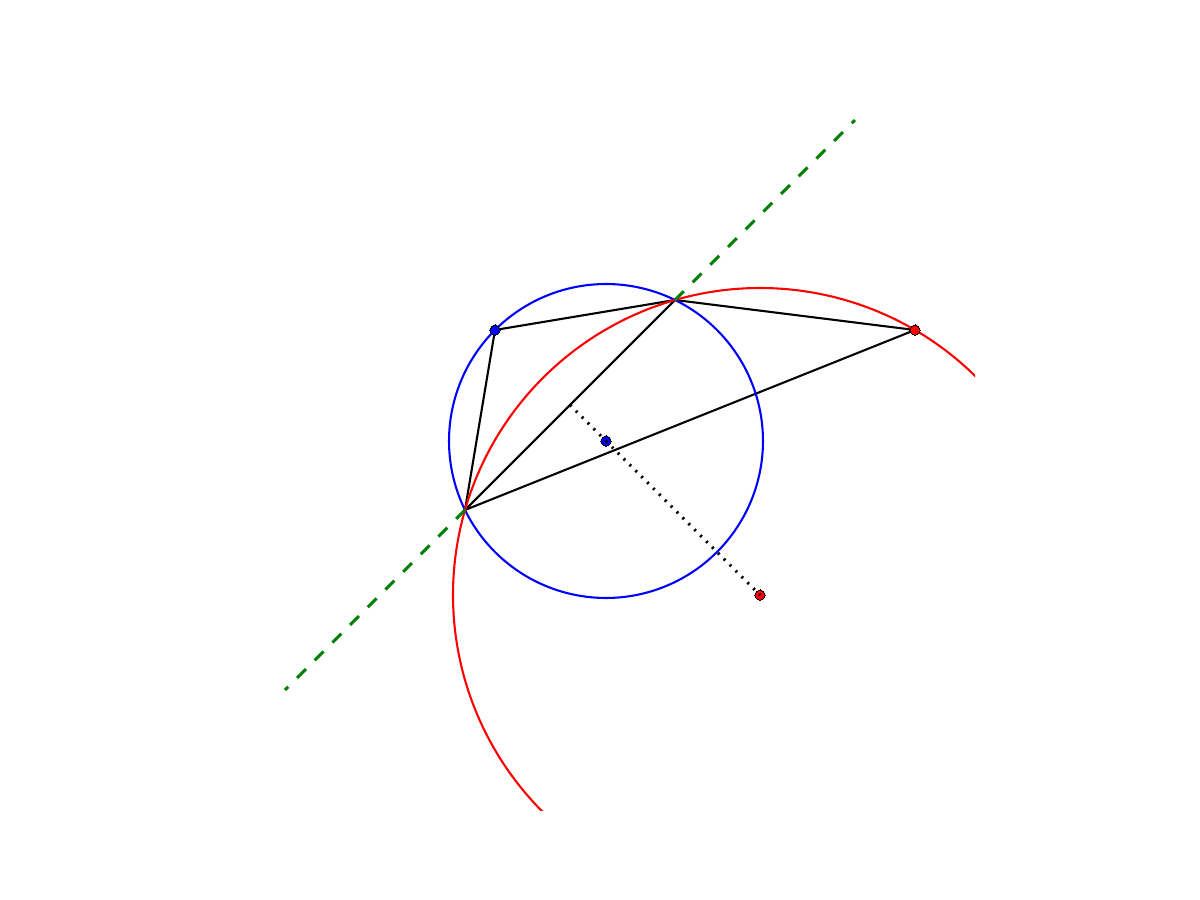},
    (0.23, 0.72) * {L},
    (0.93, 0.72) * {R},
    % (0.38, 0.56) * {c_{\tau}},
    (0.44, 0.47) * {c_{\lambda}},
    (0.70, 0.07) * {c_{\rho}},
    (0.54, 0.71) * {\tau},
  \end{cxyoverpic}
  \caption{For a Delaunay pair the ordering of the circumcenters is
    the same as that of the top dimensional simplices. See
    Lemma~\ref{lmm:crcmcntrsdlnypr}.}
  \label{fig:crcmcntrordr}
\end{figure}

The above lemma can now be used to show easily that the codimension 1
duals always have positive net length. This is the content of the next
result.

\begin{thm}[Codimension 1]
  \label{thm:cdmnsn12d}
  Let $\tau$ be a codimension 1 shared face of two $n$-dimensional
  simplices embedded in $\R^N$, $N \geq n$ forming a Delaunay
  pair. Then the signed length $\dual \tau$ is positive.
\end{thm}
\begin{pf}
  When $N = n$, the results directly follows from
  Lemma~\ref{lmm:crcmcntrsdlnypr} since the circumcenters are in the
  correct order. For $N > n$, we can isometrically embed the simplices
  in $\R^n$ in which case, the circumcenters are again in the correct
  order and the result follows.
\end{pf}

In the $N > n$ case, the signs of the elementary dual edges of $\dual
\tau$ are assigned in the affine spaces of the corresponding
$n$-dimensional simplices. For example, consider a pair of triangles
embedded in $\R^3$ and meeting at a shared edge at an angle other than
$\pi$. In this case, the signed length of the dual edge of the shared
edge is determined as the sum of the two elementary dual edges which
are measured in the planes of the two triangles individually. 

\subsection{Dual of a Vertex in Triangle Mesh Surface}
 Now we show that the area of the dual of an internal vertex in a
pairwise Delaunay triangle mesh is always positive. We prove this
below by showing that the net dual area corresponding to a pair of
triangles is positive.

\begin{thm}
  \label{thm:cdmnsn13d}
  Let $\tau$ be an internal vertex in a pairwise Delaunay triangle
  mesh embedded in $\R^N$, $N = 2, 3$. Then the signed area of $\dual
  \tau$ is a positive.
\end{thm}

\begin{figure}[h]
  \centering
  \begin{tabular}{>{\centering\arraybackslash}m{\dimexpr.45\linewidth-\tabcolsep}
      >{\centering\arraybackslash}m{\dimexpr.52\linewidth-\tabcolsep}}
    \includegraphics[scale=0.42, trim=2.0in 1.6in 2.3in 1.6in, clip]
    {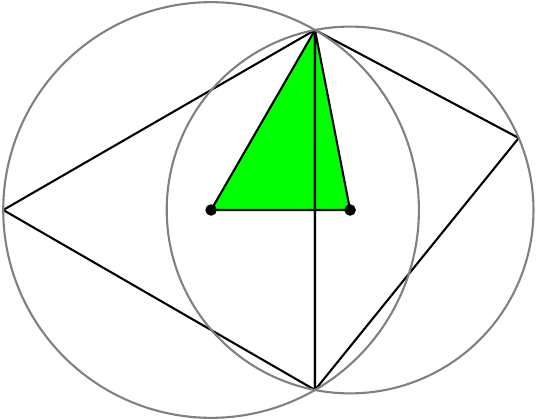} &
    \includegraphics[scale=0.4, trim=2.0in 1.4in 1.6in 0.5in, clip]
    {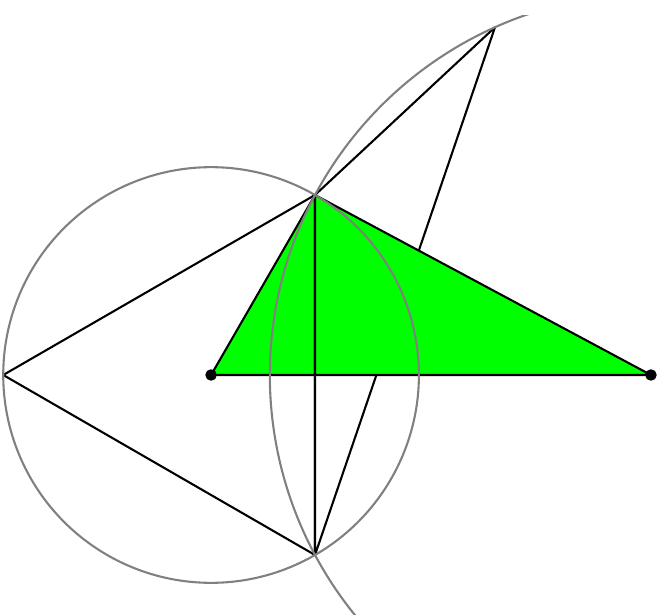} \\
    \multicolumn{2}{c}{\includegraphics[scale=0.42, trim=2.0in 1.1in 1.8in 1.0in, clip]
      {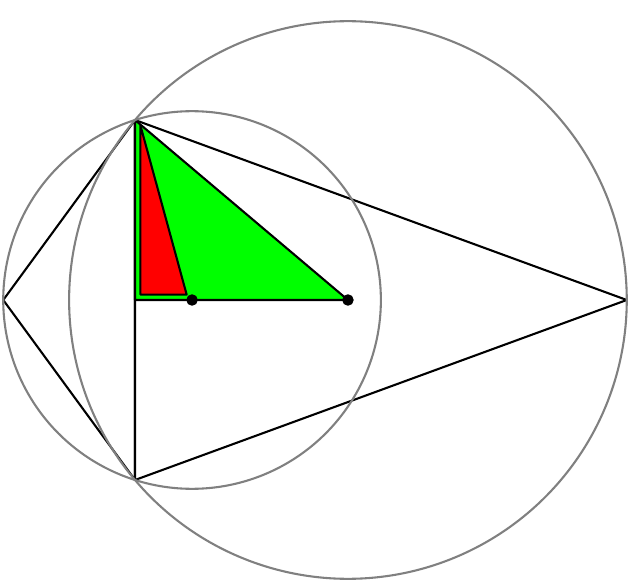}}
  \end{tabular}
  \caption{Elementary dual simplices of a vertex in a pair of
    triangles sharing an edge. The cases shown correspond to various
    positions of the circumcenters of the shared edge and the two
    triangles.}
  \label{fig:vrtxdltrngl}
\end{figure}

\begin{pf}
  $\dual \tau$ is the Voronoi cell of vertex $\tau$ in the pairwise
  Delaunay mesh. Consider a pair of triangles sharing a common edge
  incident to $\tau$ and if they are embedded in $\R^3$, isometrically
  project to $\R^2$ (i.e., treat the shared edge as a hinge, and
  flatten the pair.) The circumcenters of these two triangles are in
  correct order by Lemma~\ref{lmm:crcmcntrsdlnypr} and there are three
  possible cases as shown in Figure~\ref{fig:vrtxdltrngl}. Thus the
  net area of the two elementary dual simplices is positive when the
  signs are assigned using the rule described in
  Section~\ref{sec:sgnddlclls}. Summing over all edges containing
  $\tau$ yields the full $\dual \tau$ as a positive area.
\end{pf}

% Definition of column width with vertical centering in table below
% needed for aesthetics

\subsection{Dual of an Edge in Tetrahedral Mesh}

\begin{thm} \label{thm:edgdl}
  Let $\tau$ be an internal edge in a tetrahedral Delaunay
  triangulation embedded in $\R^3$. Then $\dual \tau$ is a simple,
  planar, convex polygon whose signed area is positive.
\end{thm}

\begin{figure}[h]
  \centering
  \begin{tabular}{c|c}
    \begin{xyoverpic}
      {(1, 1)}{scale=0.11, trim=0.5in 0.0in 0.5in 0.0in, clip}
      {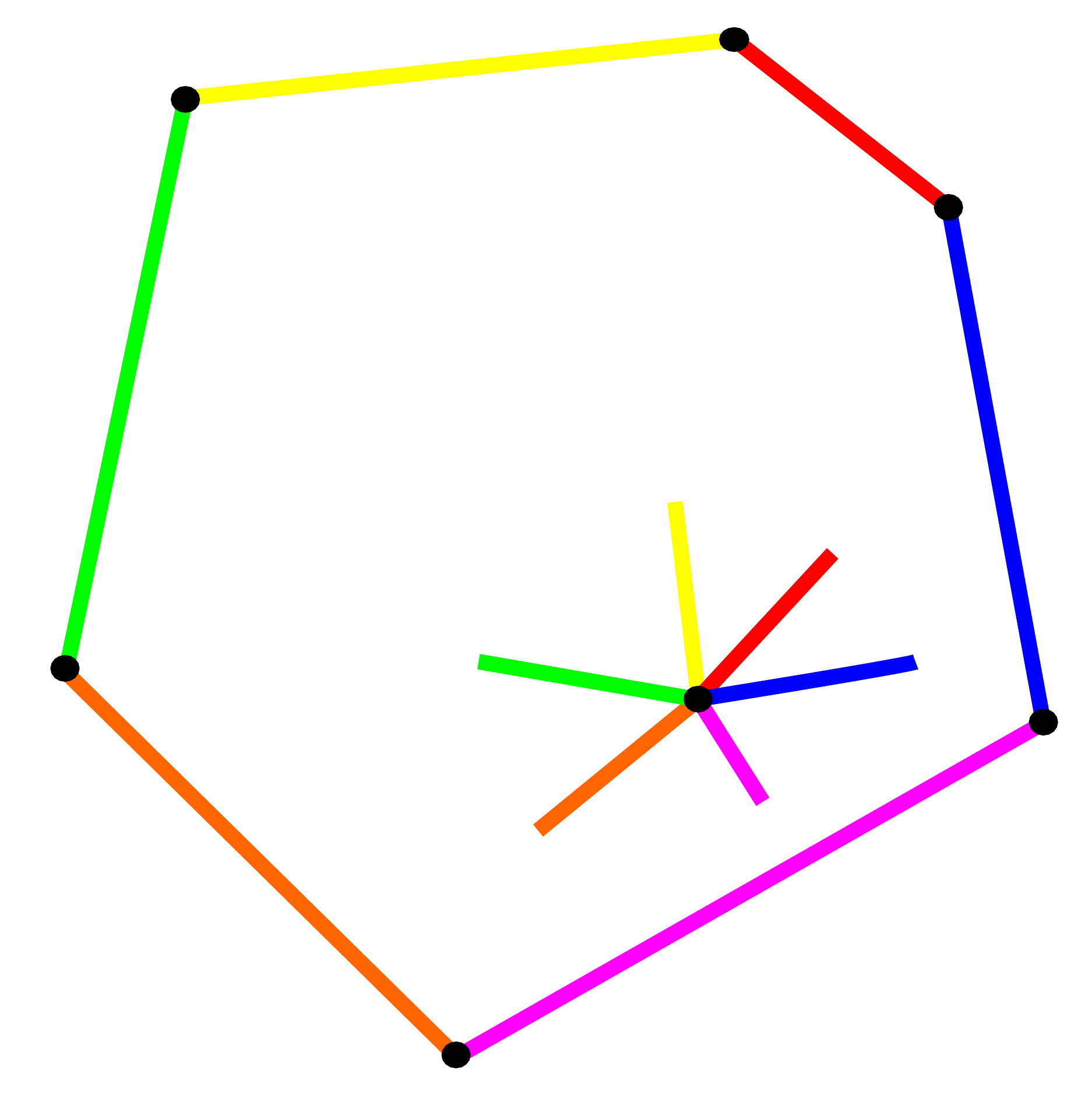},
      (0.76, 0.42) * {a},
      (0.68, 0.46) * {b},
      (0.58, 0.42) * {c},
      (0.51, 0.33) * {d},
      (0.63, 0.30) * {e},
      (0.74, 0.32) * {f},
      (0.98, 0.80) * {c_a},
      (0.76, 0.98) * {c_b},
      (0.07, 0.92) * {c_c},
      (-0.04, 0.36) * {c_d},
      (0.44, -0.02) * {c_e},
      (1.07, 0.32) * {c_f},
    \end{xyoverpic} &
    \begin{xyoverpic}
      {(1, 1)}{scale=0.125, trim=0.0in 0.0in 0.0in 0.0in, clip}
      {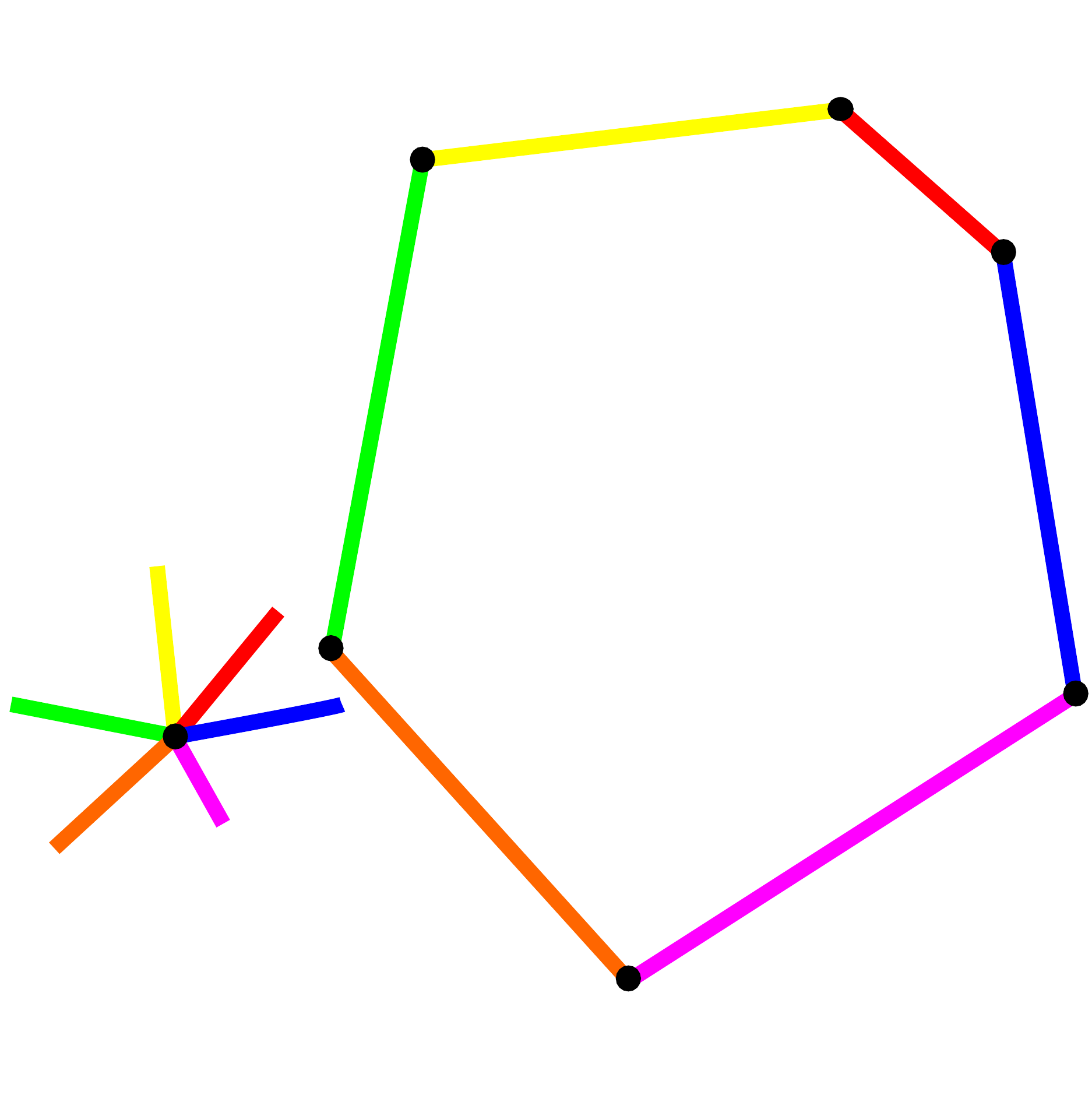}
      (0.25, 0.38) * {a},
      (0.18, 0.42) * {b},
      (0.11, 0.38) * {c},
      (0.07, 0.30) * {d},
      (0.16, 0.24) * {e},
      (0.24, 0.29) * {f},
      (0.98, 0.76) * {c_a},
      (0.83, 0.91) * {c_b},
      (0.33, 0.85) * {c_c},
      (0.37, 0.41) * {c_d},
      (0.59, 0.05) * {c_e},
      (1.04, 0.34) * {c_f},
  \end{xyoverpic}
\end{tabular}
\caption{An internal edge $\tau$ of a tetrahedral mesh may or may not
  intersect $\dual \tau$. The views here are along $\tau$ which
  appears as a point. The short lines are half-planes of the triangles
  incident to $\tau$. The tetrahedra are labeled $a$, $b$, etc. Each
  boundary edge of $\dual \tau$ corresponds to the triangle indicated
  by the coloring. The half planes could potentially be a reflection
  about $\tau$ but that is impossible in a Delaunay mesh due to
  Lemma~\ref{lmm:crcmcntrsdlnypr}.}
 \label{fig:edgdlcnfgrtns}
\end{figure}

\begin{figure}[h]
  \centering
  \begin{tabular}{c|c}
    \begin{xyoverpic}
      {(1, 1)}{scale=0.11, trim=0.0in 0.0in 0.0in 0.0in, clip}
      {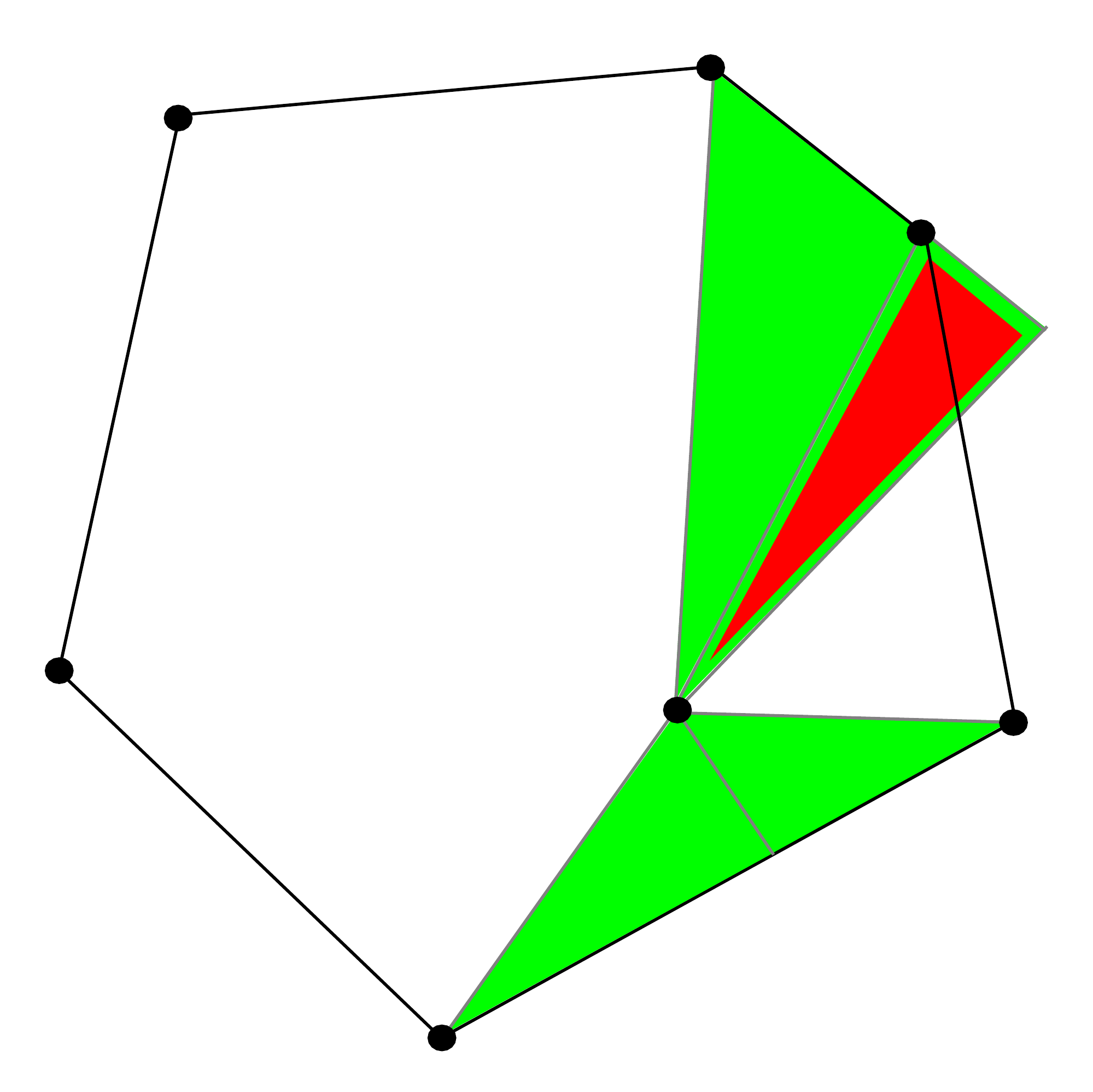},
      (0.73, 0.94) * {c_{\lambda}},
      (0.91, 0.79) * {c_{\rho}},
      (1.01, 0.70) * {c_{\sigma}},
      (0.55, 0.35) * {c_{\tau}}
    \end{xyoverpic} &
    \begin{xyoverpic}
      {(1, 1)}{scale=0.10, trim=0.0in 0.0in 0.0in 0.0in, clip}
      {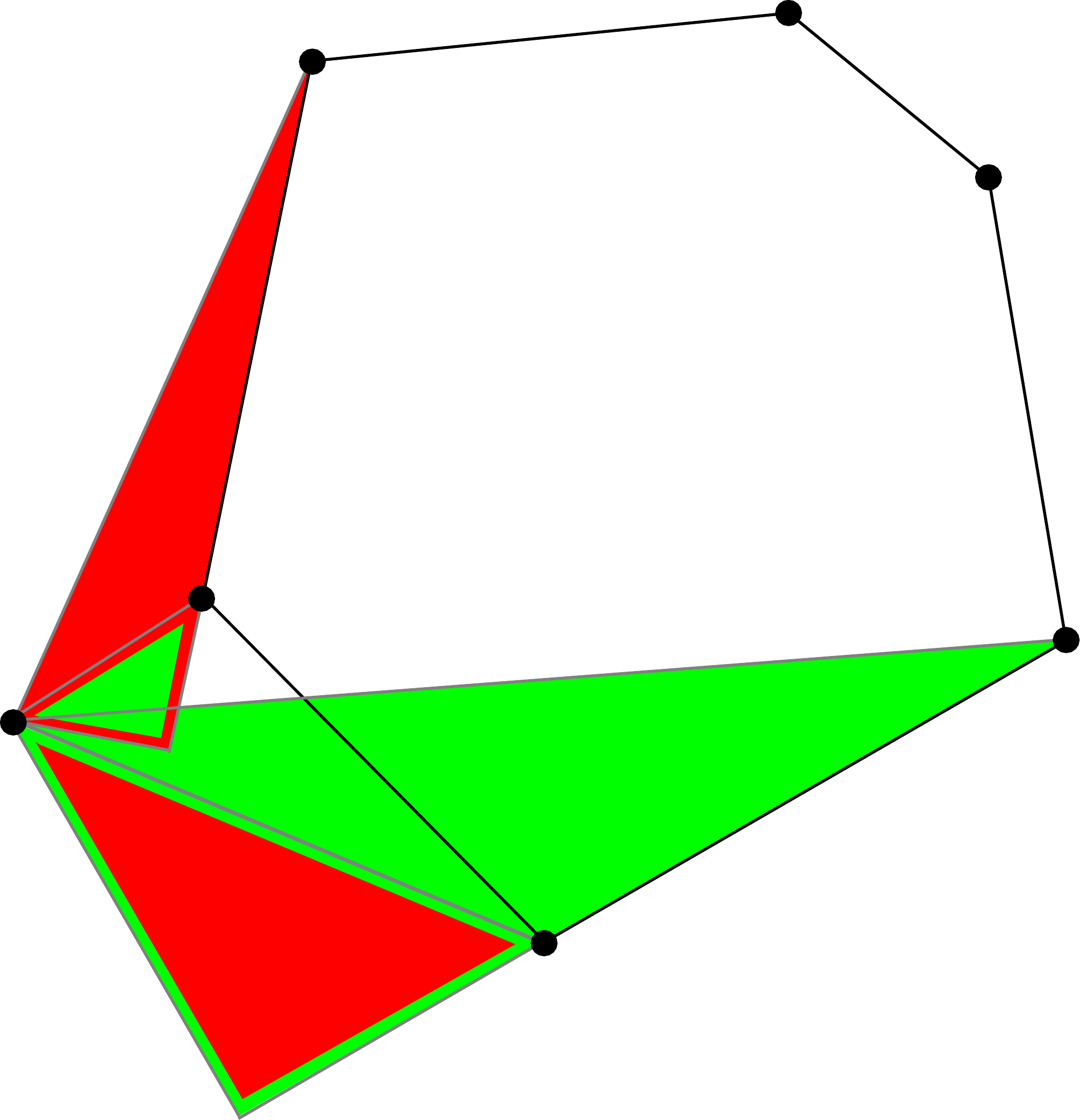},
      (0.02, 0.35) * {c_{\tau}},
      (0.32, 0.01) * {c_{\sigma}},
      (0.54, 0.14) * {c_{\rho}},
      (0.99, 0.41) * {c_{\lambda}}
    \end{xyoverpic}      
    \end{tabular}
    \caption{Representative elementary dual simplices of $\dual \tau$
      when it intersects $\tau$ (left side) and does not intersect
      $\tau$ (right side) corresponding to the two cases shown in
      Figure~\ref{fig:edgdlcnfgrtns}.}
 \label{fig:edgdlelmntrysmplcs}
\end{figure}

\begin{pf}
  $\dual \tau$ of an internal edge $\tau$ in a Delaunay triangulation
  may or may not intersect $\tau$. The vertices of $\dual \tau$ are
  circumcenters of tetrahedra incident to $\tau$ and the boundary
  edges of $\dual \tau$ are dual edges of triangles incident to
  $\tau$. Note that $\dual \tau$ is the interface between the Voronoi
  cells corresponding to the two vertices of $\tau$ and thus is a
  bounding face of both Voronoi cells. Since the Voronoi cell of a
  vertex is a convex polyhedron \cite{Edelsbrunner2006short}, $\dual \tau$
  is simple, planar and convex.

  Suppose $\tau$ intersects $\dual \tau$. Then the tetrahedra incident
  to $\tau$ and the edges of $\dual \tau$ have to be in a
  configuration shown in left part of
  Figure~\ref{fig:edgdlcnfgrtns}. A configuration in which the
  triangles incident to $\tau$ are reflected about $\tau$ is
  impossible due to Lemma~\ref{lmm:crcmcntrsdlnypr}. 

  Now, to see that the signed area of $\dual \tau$ is positive,
  consider two elementary dual simplices of $\dual \tau$ incident to a
  shared face $\sigma$ of two tetrahedra in the fan of tetrahedra
  incident to $\tau$. These two elementary dual simplices can be in
  one of the two configurations as shown in
  Figure~\ref{fig:edgdlelmntrysmplcs}. In both cases, $c_{\tau}$
  is the circumcenter of the edge $\tau$, $c_{\sigma}$ is the
  circumcenter of the shared face $\sigma$, and $c_{\rho}$ and
  $c_{\lambda}$ are the circumcenters of the two tetrahedra. Also, in
  both cases, using the sign rule of Section~\ref{sec:sgnddlclls} the
  sum of the signed areas of the elementary dual simplices is
  positive, and hence, the signed area of $\dual \tau$ composed of
  these elementary dual simplices is positive.

  Next consider the case in which $\tau$ does not intersect $\dual
  \tau$ as shown in right part of Figure~\ref{fig:edgdlcnfgrtns}. A
  boundary edge of $\dual \tau$ is called \emph{near side} if it is
  visible from the midpoint of $\tau$, otherwise, it is called a
  \emph{far side} edge. Figure~\ref{fig:edgdlelmntrysmplcs} shows the
  net dual simplices of a near side and far side boundary edge of
  $\dual \tau$. By the sign rule of Section~\ref{sec:sgnddlclls}, far
  side elementary dual simplices have a net positive signed area while
  near side elementary dual simplices have a net negative signed
  area. The negative areas of the near side dual simplices are covered
  by the positive areas of the far side dual simplices.  Thus, the sum
  of all these elementary dual simplices which is the signed area of
  $\dual \tau$ is positive.
\end{pf}

\subsection{Dual of a Vertex in Tetrahedral Mesh}

\begin{thm}
  Let $\tau$ be an internal vertex of a tetrahedra Delaunay mesh
  embedded in $\R^3$. Then the volume of $\dual \tau$ is positive.
\end{thm}
\begin{pf}
  $\dual \tau$ of a vertex $\tau$ in a Delaunay tetrahedral mesh is a
  convex polyhedron that is the Voronoi dual cell of
  $\tau$~\cite{Edelsbrunner2006short} and thus $\tau$ is inside $\dual
  \tau$. The faces of $\dual \tau$ are duals of edges incident to
  $\tau$. By Theorem~\ref{thm:edgdl} all these faces have a positive
  signed area. The direction corresponding to traversal from $\tau$ to
  an edge center always has a positive sign. Thus each pyramid formed
  by $\tau$ and a boundary face of $\star \tau$ has positive
  volume. Thus, the volume of $\star \tau$ is positive.
\end{pf}

\section{Requirements on Boundary Simplices}
\label{sec:bndry}
In the previous section we have only considered internal simplices in
a pairwise Delaunay mesh. For simplices lying in the boundary of a
domain we require an assumption to ensure positive duals.  We call a
simplex $\sigma$ \emph{one-sided} with respect to a codimension 1 face
$\tau$ if its circumcenter $c_{\sigma}$ lies in the same half space as
the apex with respect to $\tau$ in the affine space of $\sigma$.

We show below that the only assumption then needed is that a top
dimensional simplex with a codimension 1 face in the domain boundary
should be one-sided with respect to the boundary face.

Consider a pairwise Delaunay mesh of dimension $n$ embedded in $\R^N$,
$N \geq n$. Assume that $\tau$ is an $(n - 1)$-dimensional face
appearing in domain boundary and $\tau \prec \sigma^n$ such that
$\sigma^n$ is one-sided with respect to $\tau$.

\begin{thm}
  For a mesh such as above, a dual of codimension 1 faces has positive
  length. For $n = 2$ and $N = 2$ or 3, and for $n = N = 3$, duals of
  all simplices at all dimensions have positive areas or volumes.
\end{thm}

\begin{figure}[h]
  \centering
  \begin{cxyoverpic}
    {(1, 1)}{scale=0.12, trim=0.0in 0.0in 0.0in 0.0in, clip}
    {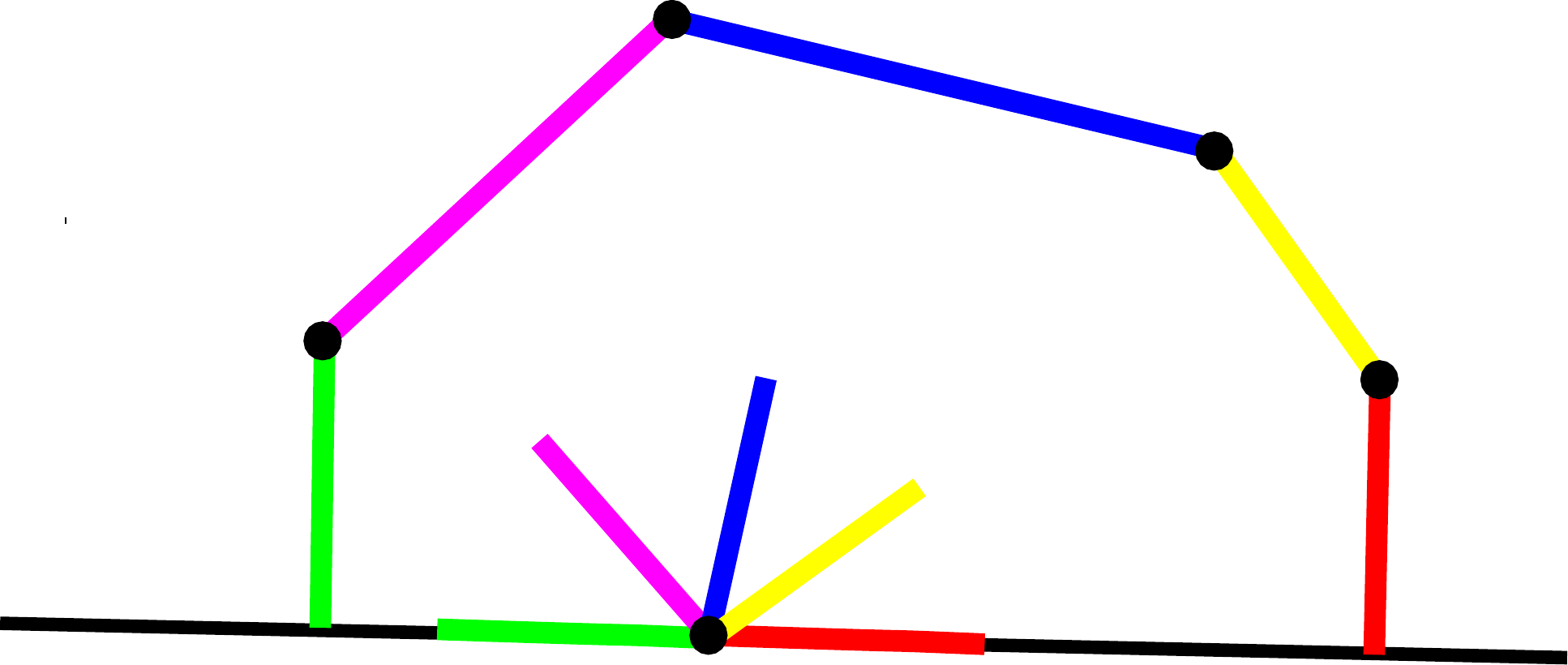},
    (0.45, -0.07) * {\tau}
  \end{cxyoverpic}
  \caption{Dual of an edge $\tau$ lying in the boundary of a Delaunay
    tetrahedral mesh. The meaning of colors and small lines is as in
    Figure~\ref{fig:edgdlcnfgrtns}.}
  \label{fig:edgdlbndry}
\end{figure}

\begin{pf}
  The codimension 1 dual of $\tau$ in all cases has positive length
  using our sign rule since $\sigma^n$ is one-sided with respect to
  $\tau$. As a result, for a surface triangle mesh, that is $n = 2$
  and $N = 2$ or 3, it easily follows from our sign rule in
  Section~\ref{sec:sgnddlclls} that the dual of a vertex on the
  boundary also has a positive area.

  For $n = N = 3$, one configuration for the dual of an edge $\tau$
  incident to the boundary is shown in Figure~\ref{fig:edgdlbndry}. In
  this figure, the plane containing the codimension 1 faces incident
  to $\tau$ are shown as short line segments, and the coloring of
  boundary edges of $\star \tau$ show the corresponding codimension 1
  face they are dual to. The other configuration in which the planes
  containing the faces incident to $\tau$ are mirror images of ones
  shown is not possible since then the circumcenters of tetrahedra
  will not be in the correct order as in
  Lemma~\ref{lmm:crcmcntrsdlnypr}. Thus, by our sign rule, all
  elementary dual simplices of $\star \tau$ are positive and hence the
  signed area of $\star \tau$ is positive. Finally, it follows from
  our sign rule that the dual of a vertex on the boundary is also
  positive since each of the elementary dual pyramids will have a
  positive volume.
\end{pf}

\section{Conclusions and Outlook}
For planar triangle meshes and for tetrahedral meshes
in three dimensional space the condition of being pairwise Delaunay is
equivalent to being Delaunay. Thus most commercial and freely
available meshing software can generate such meshes. In our
experience, several codes for planar meshing also generate meshes for
which the one-sidedness condition on the boundary is satisfied. For
example, the popular meshing code called Triangle has an option for
conforming Delaunay triangulation which generates Delaunay meshes with
one-sided boundary triangles. For tetrahedral meshes with acute input
angles this property may be harder to achieve. In general however,
algorithms for creating tetrahedral meshes with one-sided boundary
tetrahedra do
exist~\cite{Edelsbrunner2003short,Chaine2003short,GiJo2008short}.
Note that one-sidedness is equivalent to an ``oriented'' Gabriel
property (using diametral half-balls) for the boundary faces.

The pairwise Delaunay condition also appears to be more natural for
DEC than other conditions that are used in place of Delaunay in the
case of surfaces. For example, some researchers require that the
equatorial balls of triangles not contain another vertex. This
disqualifies surfaces with many folds or sharp turns. Another
alternative is to define intrinsic Delaunay condition based on
geodesics on the triangle mesh but algorithms for such surfaces can be
complicated to implement. Yet another alternative is to use
Hodge-optimized triangulations~\cite{MuMeDeDe2011short}. But creation of
these requires an additional optimization step. On the other hand
Hodge-optimized triangulation is a very interesting generalization of
Voronoi-Delaunay duality with many applications.

The invention of algorithms that generate pairwise Delaunay surface
meshes is left for future work. So is the proof of our conjecture that
the case of codimension other than 1 has positive volume for general
dimension and embedding space for pairwise Delaunay meshes with
one-sided boundary simplices. Nevertheless, the practically important
cases have all been settled by this paper.

\medskip\noindent{\bf Acknowledgement:} ANH and KK were supported by
NSF Grant DMS-0645604. We thank the anonymous referees for pointing
out some important references and for their other suggestions.

\bibliographystyle{plain}
\bibliography{dlnyhdg}

\appendix

\section{Circumcenter Order}\label{appndx:crcmcntrordr}

\begin{figure}[h]
  \begin{cxyoverpic}
    {(1, 1)}{scale=0.7, trim=2.2in 1.7in 1.5in 1.1in, clip}
    {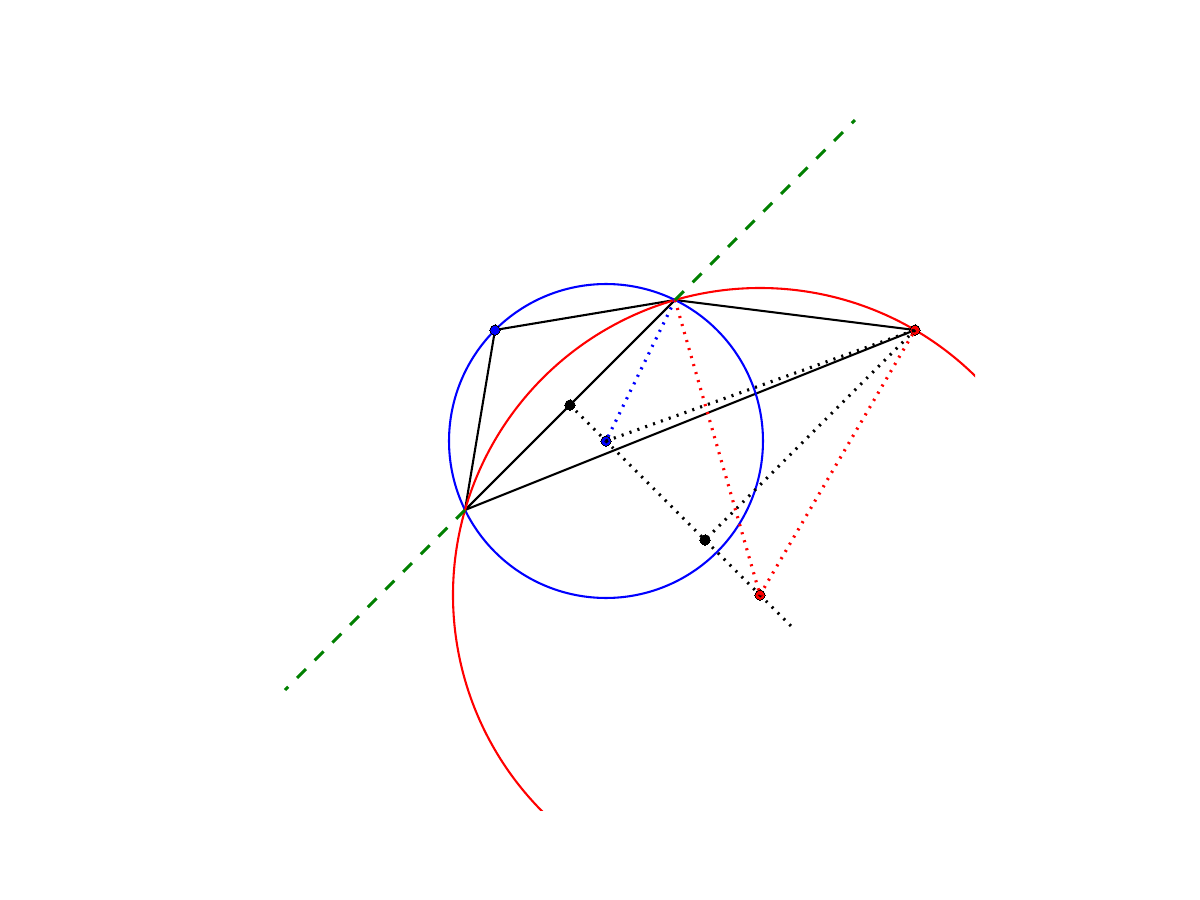},
    (0.23, 0.68) * {L},
    (0.935, 0.68) * {R},
    (0.425, 0.60) * {\tau},
    (0.34, 0.495) * {c_{\tau}},
    (0.40, 0.42) * {c_{\lambda}},
    (0.635, 0.10) * {c_{\rho}},
    (0.69, 0.03) * {\ell},
 \end{cxyoverpic}
  \caption{See proof of Lemma~\ref{lmm:crcmcntrsdlnypr2} below.}
  \label{fig:crcmcntrordr2}
\end{figure}

In fact here we prove a stronger result than
Lemma~\ref{lmm:crcmcntrsdlnypr}. We will show that the circumcenters
are in the correct order \emph{if and only if} the pair of simplices
is a non-degenerate Delaunay pair. 

\begin{lem}\label{lmm:crcmcntrsdlnypr2} 
 Let $\tau$ be an $(n - 1)$-dimensional simplex in $\R^n$. Let $L$ and
 $R$ be points separated by $\tau$. Let  $c_{\lambda}$ and $c_{\rho}$
 be the circumcenters of the $n$-dimensional simplices $\lambda = L *
 \tau$ and $\rho = R * \tau$, respectively. Then,
 $c_{\lambda}$ and $c_{\rho}$ have the same relative ordering with
 respect to $\tau$ as $L$ and $R$ if and only if $\lambda$ and $\rho$
 are a pair of non-degenerate Delaunay simplices.
\end{lem}
\begin{pf}
  Since $\lambda$ and $\rho$ are a Delaunay pair, the affine space of
  $\tau$ separates $L$ and $R$.  See
  Figure~\ref{fig:crcmcntrordr2}. Let $c_\tau$ and $r_\tau$ be the
  circumcenter and the circumradius of $\tau$, respectively. Now,
  $c_\lambda$ and $c_\rho$ lie on a line $\ell$ that passes through
  $c_\tau$ and is orthogonal to the affine space of $\tau$. Let
  $h_\lambda$ be the signed distance along $\ell$ from $c_\lambda$ to
  $c_\tau$. Similarly, let $h_\rho$ be the signed distance from
  $c_\rho$ to $c_\tau$. For now, it is sufficient that these distances
  be signed and whether the positive direction is along $L$ or $R$ is
  not important. Next, orthogonally project $R$ onto $\ell$, and let
  $r_R$ be the (positive) distance from $R$ to its projection onto
  $\ell$. Finally, let $h_R$ be the signed distance (along $\ell$)
  from the projection of $R$ onto $\ell$ to $c_\tau$.  Notice that
  $h_R$ is necessarily either negative or positive depending on the
  choice of positive direction to be either along $L$ or $R$,
  respectively.

  By elementary geometry, the squared circumradius of $\lambda$ is
  $h_\lambda^2 + r_\tau^2$ and the squared circumradius of $\rho$ is
  $h_\rho^2 + r_\tau^2$. Similarly, the squared distance from
  $c_\lambda$ to $R$ is $(h_\lambda - h_R)^2 + r_R^2$. Since $\lambda$
  and $\rho$ form a Delaunay pair, $R$ lies outside the circumsphere of
  $\lambda$. Thus, the squared circumradius of $\lambda$ is less than
  the squared distance from $c_\lambda$ to $R$:
  \begin{alignat*}{2}
    && h_\lambda^2 + r_\tau^2 & < (h_\lambda - h_R)^2 + r_R^2 \, , \\
    \Rightarrow && r_\tau^2 & < r_R^2 + h_R^2 - 2\, h_R \, h_\lambda \, .
  \end{alignat*}
  Also, since $R$ lies on the circumsphere of $\rho$, the distance
  from $c_\rho$ to $R$ is the same as the distance from $c_\rho$ to a
  vertex of $\tau$. Thus, we have:
  \begin{alignat*}{2}
    && h_\rho^2 + r_\tau^2 & = (h_\rho - h_R)^2 + r_R^2 \, , \\
    \Rightarrow && r_\tau^2 & = r_R^2 + h_R^2 - 2 \, h_R \, h_\rho \, .
  \end{alignat*}
  Using this in the previous inequality, we obtain:
  \begin{alignat*}{2}
    && r_R^2 + h_R^2 - 2 \, h_R \, h_\rho & < r_R^2 + h_R^2 - 2 \, h_R \,
    h_\lambda \, , \\
    \Rightarrow && h_R \, h_\rho & > h_R \, h_\lambda \, .
  \end{alignat*}
  Finally, we choose a coordinate direction along $\ell$ to fix signs in
  the signed distances along $\ell$.  If we choose the direction towards
  the half space containing $R$ to be positive, $h_R$ is positive. (We
  will call this the positive $R$-direction.) As a result, the last
  inequality above simplifies to $h_\rho > h_\lambda$. This means that
  $h_\rho$ is larger along the positive $R$-direction.  If we choose the
  direction along $L$ to be positive, $h_R$ is negative and we obtain
  $h_\rho < h_\lambda$.  In this case, $h_\lambda$ is larger along the
  positive $L$-direction.

  Conversely, if $\lambda$ and $\rho$ are not a Delaunay pair, then
  the distance from $c_\lambda$ to $R$ is less than the circumradius
  of $\lambda$.  Thus, all inequalities will reverse directions and
  therefore the circumradii will be in the wrong order.
\end{pf}

% \section{Main correction}
\section{Corrigendum}
\label{sec:corrigendum}
\vspace{-0.6em}

In this work (which appeared, minus the Appendix, in
\cite{HiKaVa2013}), we established positivity of entries of the
diagonal Hodge star matrix assembled from signed elementary duals for
meshes that are pairwise Delaunay, non-degenerate, and with a
one-sidedness property for boundary simplices. There are no errors in
these mathematical statements (lemmas and theorems)
in~\cite{HiKaVa2013}. Further, in~\cite{HiKaVa2013} we did not provide
any mathematical statements for meshes that violate the pairwise
Delaunay condition and/or meshes that violate the one-sidedness
condition. However in \cite{HiKaVa2013}, a numerical computation of
solution of scalar Poisson's equation was shown for such meshes. These
were shown in columns 3 and 4 of Figure 1 in~\cite{HiKaVa2013}.

We recently discovered errors in some computer programs that were used
for the numerical experiments corresponding to columns 3 and 4 of
Figure~1 in~\cite{HiKaVa2013}. Those figures in the published paper
seemed to suggest that meshes which violate the pairwise Delaunay
condition or one-sidedness condition may need to be avoided. After
correcting the errors in the programs we noticed that DEC appears to
produce the correct solution in the cases shown even for such
meshes. In light of this, we have deleted the following sentence from
the Introduction in this version as in \cite{HiKaVa2013}: ``That
figure also shows the importance of the Delaunay property and of our
boundary assumptions and the success of the signed dual volumes for
such meshes that we describe in this paper.''. We have also updated
Figure~\ref{fig:poissons}, in particular columns 3 and 4. The figures
in columns 1 and 2 of Figure~\ref{fig:poissons} in \cite{HiKaVa2013}
are unchanged by this correction -- those figures in \cite{HiKaVa2013}
(as well as the previous arXiv v3) are correct.

\begin{figure}[h]
  \centering
  \includegraphics[scale=0.5]
  {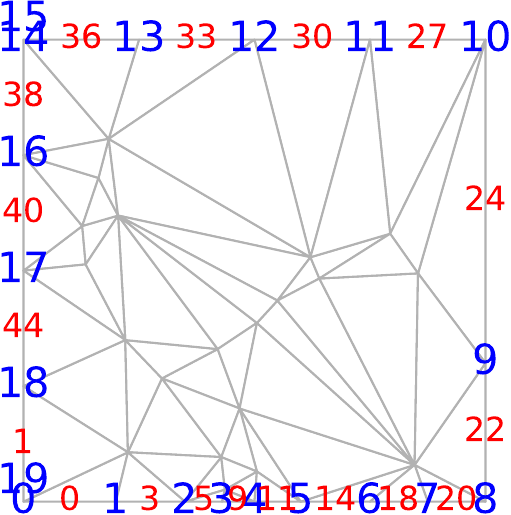}
  \caption{The cause of the programming error was an incorrect
    mesh. Notice the overlapping vertices on top left and bottom. This
    led to an incorrect assignment of boundary conditions for columns
    3 and 4 of Figure~1 in \cite{HiKaVa2013} as explained in the
    text.}
  \label{fig:bdbndryindcs}
\end{figure}

We now describe our programming error. We traced the error to the
generation and processing of mesh used for Figure~\ref{fig:poissons},
column 3 of \cite{HiKaVa2013}. The vertex and edge numberings for this
mesh are shown in Figure~\ref{fig:bdbndryindcs}. (For clarity, only
numbering of vertices and edges on the boundary are shown.)  Notice
that vertices numbered $0$ and $19$ overlap (bottom left corner) and
vertices $14$ and $15$ overlap (top left corner). This led to an
incorrect assignment of boundary conditions. (The boundary condition
for the problem is inflow through left boundary, outflow through
right, and no flow across the top and bottom boundaries.)  As a
result, the edge numbered $22$ on the right was assigned an outflow
velocity of $0$ and edge $27$ was assigned an outflow velocity of $1$
pointing to the right. Due to normal to edge $27$ pointing upwards
this resulted in $0$ flux through edge $27$ (which is
correct). However, edge $22$ was assigned an incorrect flux of
$0$. All outgoing flux was thus assigned to edge $24$ on the right.

For column 4 in Figure~\ref{fig:poissons} of \cite{HiKaVa2013}, the
non-Delaunay mesh was obtained by starting from the erroneous mesh of
column 3. In particular, one of the triangles of the mesh from column
3 was subdivided to yield a non-Delaunay pair. Consequently, the
non-Delaunay mesh inherited the edge indexing problems of the mesh
from column 3. In addition, due to modification of edge markers during
subdivision, some internal edges were misidentified as being boundary
edges. This again led to an incorrect assignment of boundary
conditions.

\end{document}